\input{psfig.sty}
\documentclass[12pt,preprint]{aastex}
\shorttitle{The Ultra-luminous M81 X-9 source: 20 years variability and 
spectral states}
\shortauthors{La Parola et al.}

\begin{document}
\title{The Ultra-luminous M81 X-9 source: 20 years variability and spectral 
states}
\author{ V. La Parola, G. Peres} 
\affil{Dipartimento di Scienze Fisiche ed Astronomiche, Sezione di Astronomia, 
Piazza del Parlamento 1, 90134 Palermo, Italy -- laparola@astropa.unipa.it;
peres@astropa.unipa.it}
\author{G. Fabbiano, D.W. Kim} 
\affil{Harvard Smithsonian Center for Astrophysics, 60 Garden St.,
02138 Cambridge, MA -- pepi@head-cfa.harvard.edu; kim@head-cfa.harvard.edu} 
\author{F. Bocchino} 
\affil{Osservatorio Astronomico "G.S. Vaiana", Piazza del Parlamento 1, 90134 
Palermo, Italy; Astrophysics Division, ESA/ESTEC, Noordwijk, The Netherlands 
-- fbocchin@estsa2.estec.esa.nl} 
 
\begin{abstract}
The source X-9 was discovered with the {\it Einstein Observatory} 
in the field of M81, and is located in the dwarf galaxy Holmberg IX. X-9 has a 
0.2-4.0 keV luminosity of $\sim 5.5\times 10^{39}$ ergs~s$^{-1}$, if it
is at the same distance as Holmberg IX (3.4 Mpc). This luminosity is above the
Eddington luminosity of a 1~$M_{\odot}$ compact accreting object.
Past hypotheses on the nature of this Super-Eddington source 
included a SNR or supershell, an accreting compact object and a 
background QSO. To shed light on the nature of this source, we have obtained 
and analyzed archival data, including the {\it Einstein} data, 23 ROSAT 
observations, Beppo-SAX and ASCA 
pointings. Our analysis reveals that most of the emission of X-9 arises from a 
point-like highly-variable source, and that lower luminosity extended 
emission may be associated with it. The spectrum of this source changes between 
low and high intensity states, in a way reminiscent of the spectra of galactic 
Black Hole candidates. Our result strongly suggest that X-9 is not a background 
QSO, but a bonafide `Super-Eddington' source in Ho IX, a dwarf companion of 
M81. 
\end{abstract}

\section{INTRODUCTION}
\label{intro}
M81 is a nearby early type spiral galaxy well studied in all
bands, from radio \citep[e.g.][]{beck, biet} to
optical and UV \citep[e.g.][]{ho} to X-rays, where it has been observed
with all the major satellites. Einstein observations \citep{fab88}
showed the presence of 10 individual X-ray sources in M81,
and 3 sources that appear
outside the optical contour of the galaxy. Among the latter ones, 
X-9 is the brightest X-ray source within 20' from the M81
nucleus, with roughly half the flux of the M81 nucleus
itself. If it is at the same distance of M81 \citep[3.6 Mpc,][]{fre}
the {\it Einstein} luminosity of X-9 is $\sim 3\times 10^{39}$ erg~s$^{-1}$, 
well above the Eddington luminosity for 1$M_{\odot}$ accretor in a binary 
system ($\sim 1.3 \times 10^{38}$ erg~s$^{-1}$). 
X-9 appears on many ROSAT observations of the M81/M82 region, and is
clearly visible in the ASCA-GIS field of an M81-centered observation
\citep{ishi}. 

\begin{figure}
\plotone{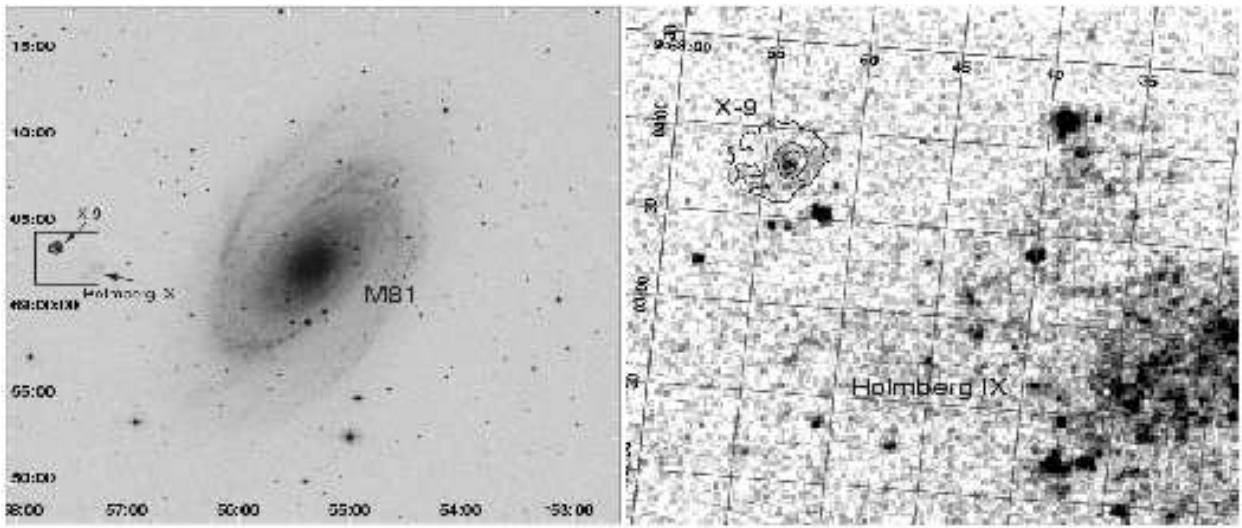}
\caption{{\bf Left panel}: contour plots of HRI count rates (smoothed with a
Gaussian filter with 2 pixel radius), superimposed to
the optical image of M81 field. X-9 is marked by an arrow. {\bf Right
panel}: Enlargement of the region within the box in the upper panel shows
both the position of X-9
with respect to Holmberg IX and the faint optical source coincident with the
X-ray emission centroid. \label{fov}}
\end{figure}

X-9 is optically coincident with the H~{\sc II} region MH9/10
\citep{mh} of the Ho IX dwarf galaxy \citep[D=3.4 Mpc,]{geo,hill}, a
member of the M81 group. \citet{miller} points out that the 
pointlike blue object reported by \citet{fab88} within 10" from the X-ray source
is indeed MH9/10, and reports an additional faint blue continuum
object in the vicinities that he identifies as an OB association.
A shock-heated superbubble with a 250~pc diameter was also found in this 
general area by \citet{miller}, who suggested a tentative physical association 
between X-9 and this extended object. If X-9 is instead a compact X-ray source, 
its super-Eddington X-ray luminosity rules out
spherical accretion onto a white dwarf of neutron star, leaving 
a black-hole binary as a possible candidate \citep[e.g.][]{fab95}. 
Another possibility, advanced by \citet{ishi} and recently explored
by \citet{ezoe}, is
that X-9 is a background quasar. Figure~\ref{fov} shows the position of X-9 with
respect to M81 and Holmberg IX.

In this paper we present a comprehensive study of the X-ray behaviour
of M81~X-9, by using all the available X-ray data, with a coverage extending 
from 1979 to 1999. Our analysis suggests that while a faint extended source 
may be indeed present, most of the X-ray emission originates from a highly
variable point-like source, with the spectral variability signature of
a black hole binary.

In Section~\ref{data} of this paper we present the data and their first 
reduction; in Section~\ref{an} we report the results of the data analysis,
including spatial structure, variability, spectral analysis. 
In Sections~\ref{disc}
and~\ref{final} we discuss our results and draw our conclusions.

\section{OBSERVATIONS AND DATA REDUCTION} 
\label{data}
Table~\ref{log} summarizes the log of all the X-ray observations of X-9. 
X-9 was observed with both {\it Einstein } HRI and IPC 
\citep{giac} in 1979 \citep{fab88}.
It was then observed 23 times with ROSAT in a period of seven years from 1991 
to 1998: 12 times with the PSPC 
\citep[Position Sensitive Proportional Counter,][]{pfef} and 11 times with the 
HRI \citep[High Resolution Imager][]{david}.
It was also observed in 1998 with BeppoSAX \citep{boella}, and in 1999 
with ASCA \citep{tanaka}.
We therefore have observations spanning 20 years, with good coverage in the 
second half of this period.
With this data, we can: (1) study the temporal behaviour of X-9 (using all
instruments for light curves and fluxes); (2) study its spectral 
characteristics and how they vary with time, using  ROSAT PSPC data, as well as 
the wider-band higher-resolution BeppoSAX 
and ASCA spectra; (3) re-examine the spatial properties suggested by
\citet{miller}, using the coadded good-resolution ($\sim 7"$) ROSAT HRI data.

\subsection{Einstein Data}
The {\it Einstein} satellite observed M81 and the 
surrounding region in 1979. Three pointings were made: one with the IPC (Imaging
Proportional Counter) in April and two with the HRI (High Resolution Imager) in 
May and September. 

The IPC data cover a spectral energy range spanning from 0.2 to 3.5 keV and 
have a spatial resolution of $\sim1'$. The source counts were extracted 
from a circular region of 180" centered on the centroid position \citep{fab88}
yielding a count rate of $0.066\pm0.003$ count~s$^{-1}$.  A low resolution 
spectrum can be in principle obtained from IPC data,
but the source falls in a region of the detector where the spectral calibration
is highly uncertain. For this reason we only used these data to derive an
estimate of the source flux, using the spectral models suggested by the other
available spectral data (see Section~\ref{spec} and Table~\ref{models} for 
details).

The Einstein HRI has a spatial resolution of $\sim3"$, and no spectral 
resolution. Photons from M81 were collected during two different observing 
period, then merged in the same file. However, the two relevant pointings are 
slightly offset. For this reason X-9 is in the HRI field only in the September 
data. The source counts (that 
thus refer just to September, as well as the exposure time in Table~\ref{log}) 
were extracted from a 18'' radius circle centered 
on the source centroid and the count rate is $\sim 0.12$ counts~s$^{-1}$.
For full details on source counts and background extraction and events 
corrections we refer to \citet{fab88}. 

\subsection{ROSAT Data}
ROSAT data were processed using the IRAF(v. 2.11)/PROS (v. 2.5) software system
\citep{tody,wor}.
The twelve PSPC observations (see Table~\ref{log}) were aligned using the 
nominal pointing directions. The result was then checked by
comparing the position on the image array of three bright sources in the field 
of view. In order to achieve a better signal-to-noise ratio when determining 
the source position, the eight M81-pointed observations were coadded and the 
wavelet detection algorithm \citep{dami} used in the {\sc galpipe} pipeline 
analysis of ROSAT/PSPC data \citep{mak} was applied to
the coadded image. This algorithm is based on the application of a set of 
wavelet transforms of the image, evaluated on different radii, and allows the 
detection of both point-like and extended sources, the determination of
source centroid position, and an estimation of its count-rate and size.
The remaining four PSPC sequences (2P, 3P, 4P, 9P) were not used in this 
process because they show X-9 at a large off axis angle, so that the source 
image is highly degraded. The centroid coordinates were found to be 9h 57m 54s 
RA and $69^{\circ}03'46"$ Dec. 
The average count rate of the source is $0.196 \pm 0.002$ cnt~s$^{-1}$, with a 
total 
exposure time of 146 ksec. The source radius (3.25') was chosen to be the point 
where the radial profile of the source (in 20" bins) become consistent with the 
value of the field background (extracted from the source-subtracted image, from 
an annulus centered on the centroid position of the nucleus of M81, with radii 
540" and 600"). The same coadded image was used to extract the spectral
information (see Section~\ref{spec}). 
In order to conduct the different steps of time analysis, we examined the 
source emission in each individual observation. The radius 
for source photon 
selection depends on the position of X-9 on the detector plane, because of the 
increase of the radius of the PSF with increasing distance from the detector 
center \citep{boese}. For the off-axis observations we used the source radius 
evaluated with the {\sc galpipe} 
processing applied on each observation: 4' to 5' 
where X-9 position is between 20' and 40' away from the
detector center. We have excluded any part of the
selected region obscured by the instrumental PSPC "ribs".

The source is also visible in eleven HRI observations pointed on M81, with a 
total exposure time of 177 ksec and an average count rate of $(5.31 \pm 0.09) 
\times 10^{-2}$ cnt~s$^{-1}$. Among these, sequence rh180015n00 (2H in 
Table~\ref{log}) has been excluded from the 
analysis because we were not able to detect any source in the whole field of 
view with the detection algorithm in the IRAF/PROS task {\sc xexamine}, 
probably due to its short exposure time. This made it impossible 
to align this sequence with the other ones. The remaining observations were
aligned by comparing and aligning the position of three bright sources in the 
field of view. The task {\sc xexamine} was 
used to find the source centroid position from the longest live time observation
(1H). The X-9 source coordinates are 9h 57m 54.3s RA and $69^{\circ}03'46.4"$
Dec, well within the error box associated with the PSPC
detection. Thus the two determinations are consistent with each other and from 
now on we will use the more accurate HRI centroid coordinates. For each
observation, source counts were extracted 
from a circular region with 1.3' radius, as found with the {\sc ldetect} 
IRAF/PROS algorithm (S. Dyson 1999, private communication).

\subsection{BeppoSAX Data}
X-9 is visible in the BeppoSAX observation of M81 performed on June
5th, 1998 \citep{pelle}. The observation was processed with the SAXDAS software
version 1.6.1 applying the standard corrections and event selections.
It includes data from the Low \citep[LECS,][]{parmar}
and the Medium \citep[MECS,][]{boella} Energy Concentrator System,
which are sensitive in the 0.1--10 keV band (0.1--4.5 keV and 1.6--10.0 keV
respectively), have an angular resolution of
2' FWHM and an energy resolution $\Delta E/E=0.08$ (FWHM) at 6 keV.
The LECS and MECS exposure times are 23 and 69 ksec, respectively.

We have extracted the LECS spectrum of X-9 from a
circular extraction region with radius of 4',
which ensures that we collect at least 80\% of the photons
of this faint source. The extraction region is centered at
9h57m52.7s RA, +$69^{\circ}03'47"$ Dec (J2000).  The background was
collected from a ring with inner and outer radii of 10.5' and 13.8',
respectively, centered on the on-axis position, excluding the X-9 source
regions defined above. The spectrum was rebinned to have at least
20 counts per energy bin. We have generated the LECS response matrix
and ancillary response file using the task {\sc lemat} in the LHEASOFT
Release Version 5.0.1. We have taken peculiar care in the analysis of the MECS
data, because of the partial source obscuration by the support ribs. A small 2'
extraction radius was chosen to exclude the ribs,
and the MECS2 and MECS3 spectra were extracted 
separately. We have verified that, even with this small extraction
region, the MECS3 spectrum is still severely distorted by the rib
obscuration and therefore we have chosen to use only the MECS2
spectrum.  We have used the standard MECS response matrix and the
ancillary response file for an off-axis source located at 14' (X-9 is
at $\sim 12'$ offaxis) and an extraction radius of 2', created using
the on axis 2' extraction radius file available from the Sax Data Center
and correcting for the energy depndent vignetting effect.  MECS
background selection and rebinning are similar to those of the LECS.

\subsection{ASCA Data}
ASCA observed X-9 in 1999; all instruments were active. For this analysis, we
used the archival event files,
with the default screening and event selection criteria, were used. During the 
observation the GIS detectors \citep[Gas Imaging Spectrometer;][]{koh} 
were in the PH normal mode. The source counts were extracted from a 6' 
circular region around the source centroid and the background was subtracted 
from a
source-free circular region in the same field. The count rates for the two GIS
detectors in the (0.5-10.0) keV band are $0.217 \pm 0.003$ (GIS2) and 
$0.270 \pm 0.003$ (GIS3) count~s$^{-1}$,
for an integration time of 38 ksec. For the spectral analysis, the 
most recent publicly released response files were used for each detector.
We also estimated the expected contamination from the nucleus of M81 in the 
extraction region of X-9, by using the calibration image of Cyg X-1 provided by 
the HEASARC. We estimate that at least 7\% of the counts extracted from
the X-9 region belongs indeed to the nucleus of M81.

The two SIS \citep[Solid-state Imaging Spectrometer;][]{burke} detectors 
were observing in 1CCD BRIGHT mode. Chip 1 on SIS0 and chip 3 on SIS1 were 
active. We extracted the source counts from a 4' circular region 
around the source centroid. The contamination due to photons 
from the nucleus of M81 (which falls outside the chip) is less than 0.02\% 
(Ebisawa 2000, private communication). The background photons were collected 
from the 
source-free region of each chip in the same observation. The response matrix 
and ancillary file were produced with the FTOOLS {\sc sisrmg} and {\sc ascaarf} 
(FTOOLS v3.5), separately for each detector.
The two SIS count rates for X-9 are $0.350 \pm 0.004$ (SIS0) $0.288 \pm 0.003$
(SIS1) cnt~s$^{-1}$, for an integration time of 33 ksec. 

We then compared the SIS and GIS spectra in order to estimate the amount of 
contamination from the nucleus of M81 to the spectrum of X-9 in the GIS
data. The GIS spectrum shows a well detectable soft excess over the SIS 
spectrum. As the four instruments on board
of ASCA observe the target simultaneously, this difference cannot be due to an
intrinsic variation of the source emission: the most likely explanation is
instead the contamination from the nuclear source. In order to check this 
hypothesis we extracted the spectrum of the nucleus of M81 from the two GIS 
fields (the radius of the extraction region is 6'). This
spectrum is very well described by a power-law with $\Gamma=
1.80^{+0.02}_{-0.01}$ ($\chi^2/\nu=314/326$). We used this model to fit the
soft excess found in the GIS data when the overall spectrum is modelled with 
the best fit SIS parameters (see Section~\ref{spec} and Table~\ref{models}
for details). This excess amounts
to $\sim 20\%$ of the total photon flux, that is twice the expected
contamination. As we where not able to fully explain the difference between the
observations with the 
two instruments, we decided to exclude the GIS data set from the following
analysis, retaining only the SIS data. 

Incidentally, we note that in this observation (1999) the nucleus of M81
appears to be in a different spectral state with respect to that of the
1993 observations, reported by \citet{ishi}, where a thermal component was 
needed, in addiction to the power law, to fit the observed spectrum. 

\section{ANALYSIS AND RESULTS}
\label{an}
\subsection{Spatial Properties}
\label{image}
In order to establish the nature of X-9, it is very important to 
determine its shape and extension. Our wavelet detection analysis (see Section
2.2) suggests a radius of 3'.25 for this source in the PSPC image, which
is larger than the predicted PSF radius for the relevant off axis angle
\citep{boese} and suggests that X-9 could have an extended
component: to check this hypothesis we used the better angular resolution
HRI data. The main difficulty in this analysis is 
that X-9 is 12'.5 off axis, therefore the analytical formula for the
bi-dimentional point spread function cannot be used,
and alternative methods must be applied. We analysed the data  
from the HRI sequence 1H (that has the longest live-time) and compared the
results to those from the 
point-like calibration source HZ43, whose data are in the IRAF calibration 
database. We chose the HZ43 observation at 12' off axis which has the  
off axis angle closest to that of X-9. We did 
not use the coadded file obtained from the ten available observations of X-9, 
in order to avoid any small misalignement that could possibly
affect the coadded image. As a first step, we applied to the 1H data the aspect 
correction required for data processed before the SASS7-B
release\footnote{From http://hea-www.harvard.edu/rosat/rsdc\_www/aspfix.html}
as well as the dewobbling procedure of \citet{harris}.
We found that this correction does not appreciably affect the data. 
The source centroid position, determined using the detection algorithm
in the IRAF/PROS task {\sc xexamine}, is 9h 57m 54.3s RA and 
$69^{\circ}03'46.4"$ Dec. 

\begin{figure}
\plotone{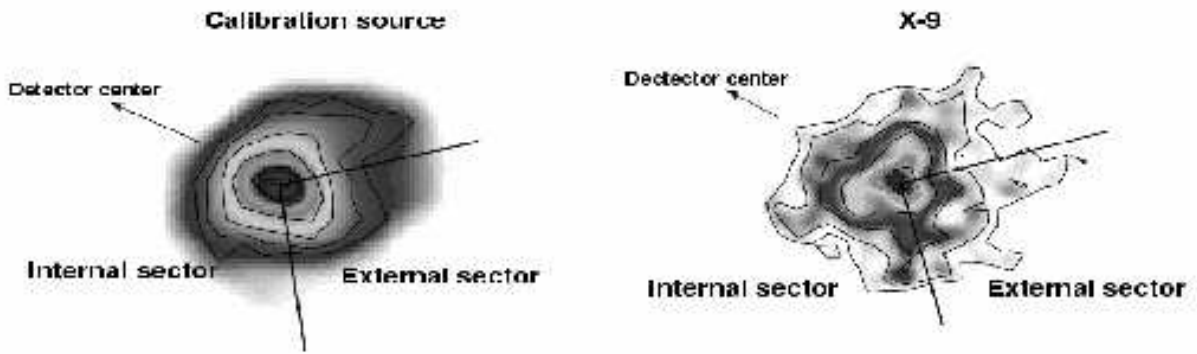}
\caption{{\bf Left}: contour plots of HZ43 calibration source counts on ROSAT 
HRI at 12' off axis, where the asymmetric structure of the PSF is evident.
{\bf Right}: contour plot of X-9 counts have been rotated to have the same 
position as HZ43 respect to the detector center. Sectors
used for the radial profile extraction are also shown, as well as the direction
of the detector center. \label{calcont}}
\end{figure}

\begin{figure}[h]
\centerline{\psfig{figure=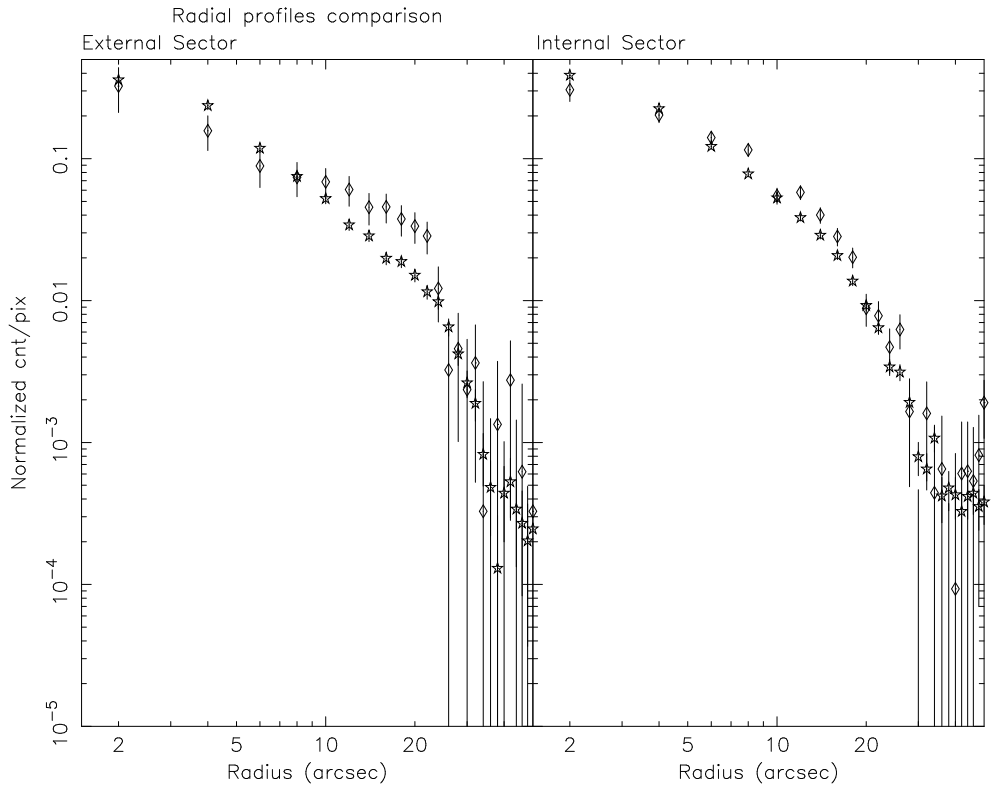,width=15cm,angle=270}}
\caption{Radial profile of the calibration source HZ43 (stars), 
compared to X-9 profile (diamonds). Left panel: external sector. 
Right panel: internal sector. Data points are normalized to the total number 
of counts
within 2'. Extraction annuli are 2" wide (1 detector pixel = 0.5"). 
\label{rprof}}

\end{figure}

The X-ray contours of the calibration HZ43 image, smoothed with IRAF {\sc 
imsmooth} with a 2 pixels Gaussian function, show that the PSF is  
distorted, with a larger width in the direction of the center of the detector
(Figure ~\ref{calcont}, left panel). The contours from the 1H image, smoothed in
the same way, show a similar distorted shape with comparable extension, but due 
to the smaller S/N ratio, the shape is less well determined (Figure 
~\ref{calcont}, right panel). To compare 
the radial profile of the two sources, we selected two angular sectors (see 
Figure~\ref{calcont}) centered on the source centroid positions: the internal 
one is in the
direction of the detector center and the external one in the opposite direction.
The relevant angles were selected on the calibration image, and then
reported on the X-9 image. 
Counts were then extracted, for each sector, from 60 annuli each 2" wide 
around the source centroid, for a total radius of 2'. In order to
make a direct comparison of the two profiles, both were normalized to the total 
encircled counts within a 2' radius. In Figure~\ref{rprof} we plot both the 
internal and the
external profiles for the two sources. In both profiles X-9 appears more 
extended than HZ43. In particular between 10" and 20" of the external
sector there is an evident excess of X-9 with respect to the calibration curve 
that is well below the error bars. This excess corresponds to about 10\% of the
total emission, i.e. a luminosity of $\sim 2.5\times 10^{38}$ for a line of sight
galactic $N_H=4.1\times 10^{20}$ cm$^{-2}$, assuming a 1 keV thermal spectrum at the 
distance of Holmberg IX. 
This has led us to the conclusion that X-9 is likely to be a composite source,
including a point like component (as supported by the time variability analysis,
see next Section) and  diffuse emission, which may be physically unrelated.

The comparison of the HRI image with the optical one from 
the POSS~II archive confirms the coincidence of the center of the brightest 
X-ray countour with the faint blue object reported by Fabbiano as the probable 
optical counterpart of X-9 and identified by \citet{miller} with the
H~{\sc II} region MH9/10 (Figure~\ref{fov}). \citet{henk} estimated the 
magnitude of this
object (that appears extended on a size of 20'') to be  $m_B=20^m.1\pm 0^m.1$. 
However, the angular resolution of our data 
is still insufficient to give strong constraints on the spatial distribution of 
X-9 and to argue on a physical association between the X-ray
emission and the optical object.

\subsection{Time Variability}
\label{time}
\begin{figure}[h]
\plotone{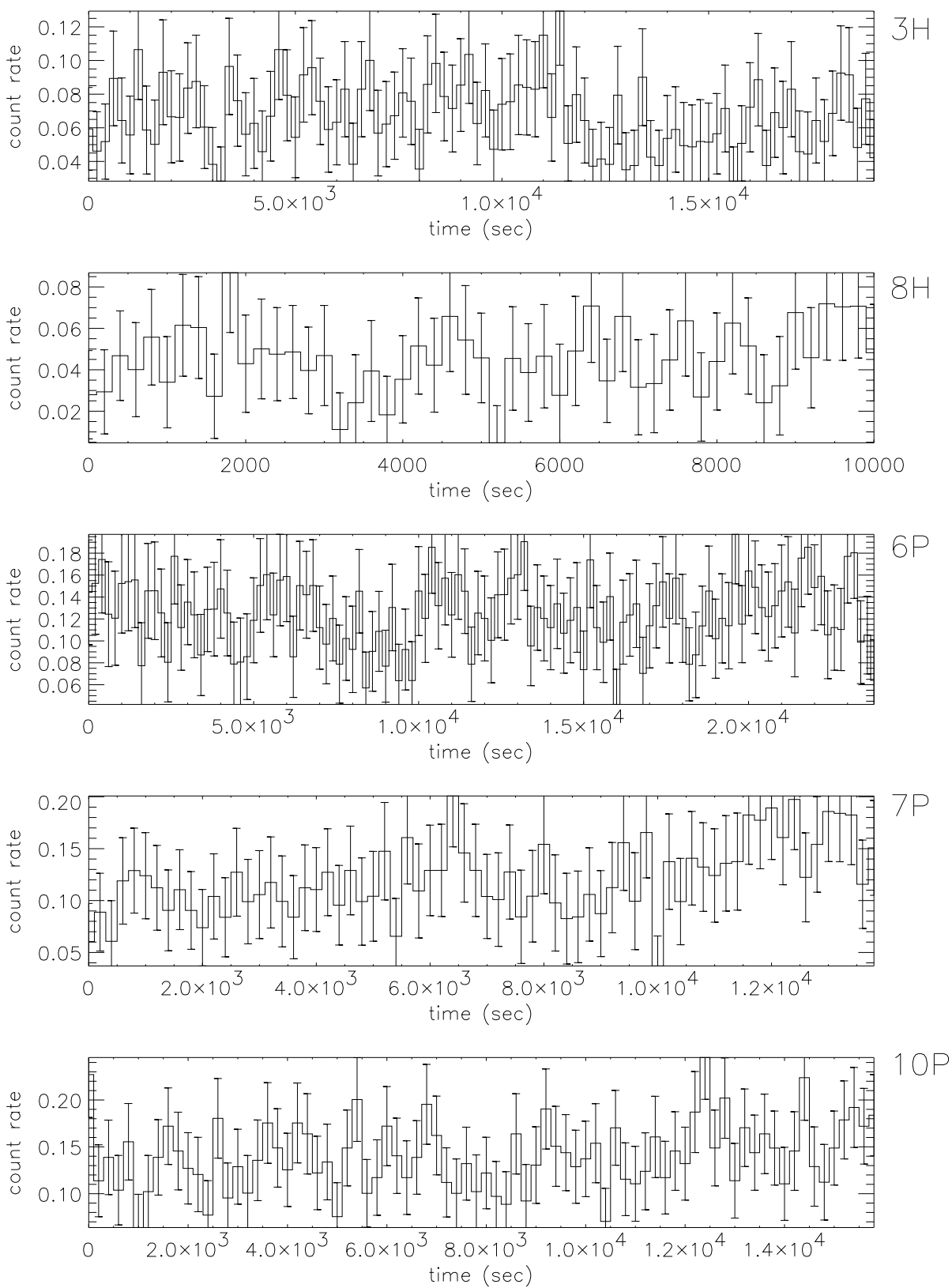}
\caption{Individual lightcurves in 200 s bins of the variable observations 
(identified by the code on the right of each plot). For clarity of presentation 
only the GTI are plotted.}
\label{ltc}
\end{figure}

We studied the temporal evolution of X-9 using {\it Einstein}, ROSAT, 
SAX and ASCA data. Source and background photons were extracted as explained in 
Section~\ref{data}. Our data cover a period of $\sim 20$ years spanning from 
April 1979 to April 1999. We searched for short scale variability within 
each observation with a Kolmogorov-Smirnov and a Cramer-von Mises 
variability test. We excluded from this analysis the four PSPC off-axis 
observations, as the position of the source on the detector makes the 
lightcurve much more affected by the strong background variations caused by the 
detector wobbling. We report the results in Table~\ref{kstest}. In four 
observations out of twenty two we detected variability with, at 
least, 90\% confidence with both tests. We performed the same tests 
on the background field in the vicinity of the source, finding that the 
background emission is variable as well, with a typical time scale of the order 
of $10^3$ sec. This time scale was estimated from the oscillations of the 
background counts around the curve of uniform distribution hypothesis, and then 
compared with that of the source (evaluated in the same way). We
found that in all cases there is no evident matching between the background and
source time scales, and we conclude that the source variability we find is 
real, and not an artifact due to background variability.
To check further the consistency of our results, we performed a different
test \citep{coll} based on $\chi^2$ statistics applied on the 
background subtracted light curves (created in 200s bins) for each observation. 
This test also checks for the variability time-scale through a pulse model. 
We thus consider to be variable
only those observations with more than 90\% chance of variability and with at
least two methods. This includes observations 6P, 7P, 10P, 3H, 11H. From the 
inspection of the individual lightcurves (Figure~\ref{ltc}) there is no 
striking evidence of either periodic or flare-like variability features. In
the observation 7P we observe a source  flux steadily increasing.

\begin{figure}[h]
\plotone{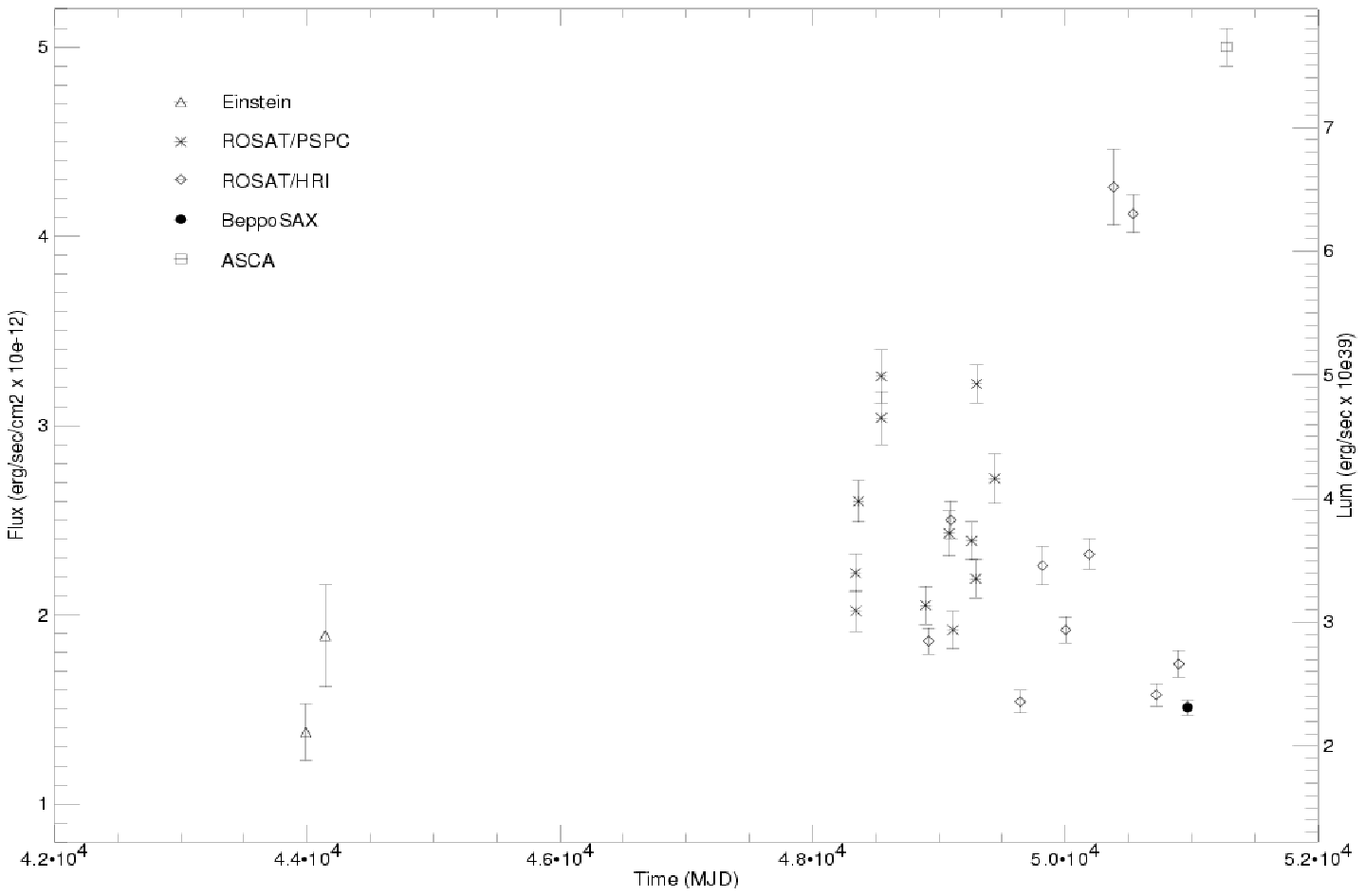}
\caption{X-9 flux between 0.5 and 2.4 keV as a function of time. See text for
details on flux calculation. The luminosity
assumes a distance of 3.4 Mpc}
\label{cntrate}
\end{figure}

In order to study long term variability, we compared data from the 
different instruments using all the available observations. Source counts
were extracted from each observation as described in Section~\ref{data} after
correcting for the appropriate exposure map. Exposure maps for ROSAT/HRI were
created using the analysis procedure tools developed by \citet{snow}.
For each PSPC observation we calculated the 
source flux in the (0.5-2.4) keV energy band using the parameters that best fit 
the PSPC spectra (see Section~\ref{spec} and
Table~\ref{models} for full details). For ROSAT/HRI and Einstein (IPC and HRI) 
observations we adopted the model
used for PSPC data, correcting for the different count rate and instrumental 
response. The appropriate best fit models were used for ASCA and SAX. All fluxes
were calculated in the (0.5-2.4) keV band.
Results are reported in Figure~\ref{cntrate}, where each data point corresponds 
to one observation. Here we note that, in spite of being
taken with different instruments, HRI and PSPC data are in
good agreement, and this is particularly evident in the period that covers 
April 1992. Moreover SAX data are consistent with the 
trend shown by the last HRI observations, while the ASCA point shows a much
higher flux. The source flux
is clearly variable: in Figure~\ref{cntrate} there is evidence of variability 
on time scales of months as well as a descending trend on longer time scale 
(few years) broken by some flare-like episodes. The data do not 
show any hint of periodical variability. A flare or a change in the physical
status of the source could be responsible for the two high luminosity episodes 
shown by the two HRI observations 8H and 9H (at $\sim 5.05 \times 10^4$ MJD) 
and by the ASCA observation.

\begin{figure}[h]
\plotone{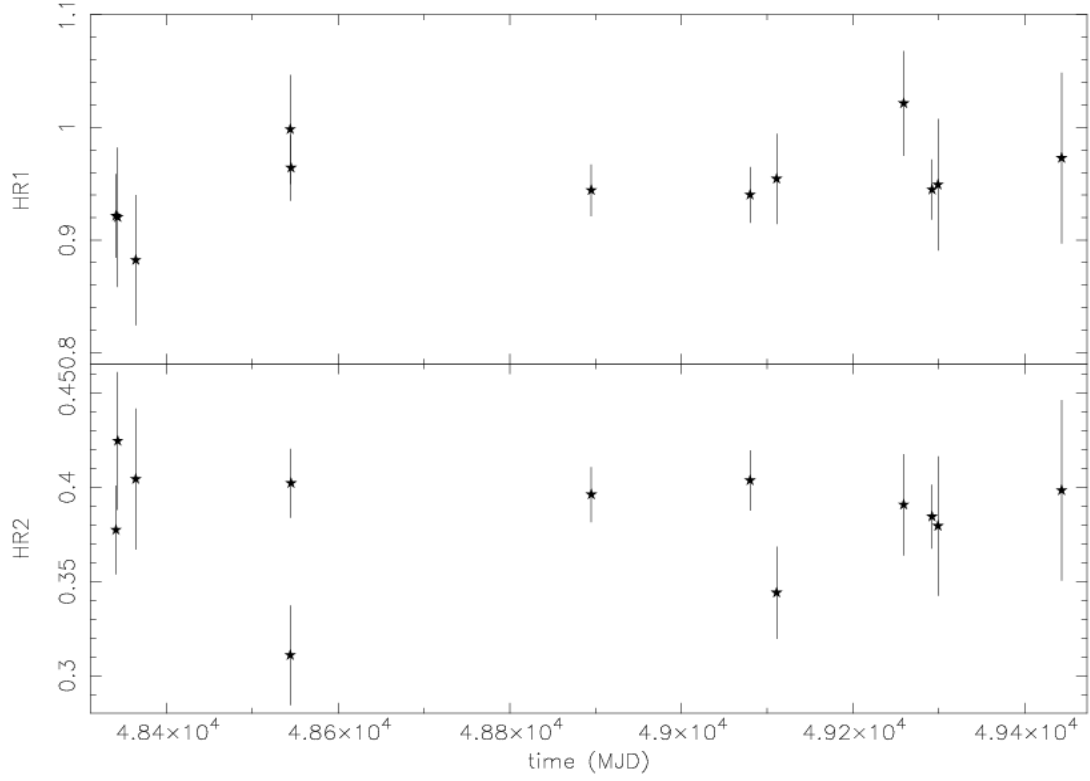}
\caption{Hardness ratios HR1 and HR2 for X-9 in M81 from 
ROSAT-PSPC data. Ratios HR1 and HR2 are defined as 
$\frac{hard-soft}{hard+soft}$, see Table 3. \label{hr}}

\end{figure}

We also searched for spectral variability in the PSPC data, by evaluating two 
different hardness ratios, both defined as 
\begin{equation}
HR = \frac{hard - soft}{hard + soft} 
\end{equation}
using for each the energy bands showed in 
Table~\ref{hrdef}. We decided to use two different spectral hardness ratios 
because the first gives a measure of the spectral hardness over the whole PSPC
band, while the second allows us to probe the spectral behaviour of the high 
energy part of the ROSAT/PSPC band. We did not
use the data from ASCA, BeppoSAX and {\it Einstein}, as the statistics of X-9
allows to obtain only a single point from each observation and we cannot 
compare points from different instruments.
We plot both hardness ratios in Figure~\ref{hr}.
We find no significant variability of the hardness ratios, with the
only exception of a single point showing a softer spectrum (observation 4P) in 
the high energy band ratio (HR2).

\subsection{Spectral Analysis} 
\label{spec}

We have performed model-fitting analysis of PSPC, BeppoSAX and ASCA spectral 
data of X-9.
In order to achieve a good S/N ratio we summed the eight ROSAT-PSPC pointed
observations of M81. PSPC Spectra were extracted from a circular region
with radius 360"; background is from an annulus centered on the source,
with radii 360" and 500" (internal and external respectively), excluding 
the slice falling on the instrumental rib.
Spectral data from the Beppo-SAX and ASCA observations were extracted as
explained in Section~\ref{data}. We have thus obtained 
different spectra over a total energy range going from 0.5 to 10.0 keV.
The analysis was performed using the XSPEC v.10 software tool \citep{arn}.

We examined several hypotheses to explain the strong X-ray emission of X-9,
including the possibility that X-9's Super-Eddington 
luminosity is the signature of an accreting black hole \citep[see][]{fab95}. 
We used the following spectral
models to fit the spectra from the different instruments:
\begin{itemize}
\item an absorbed power-law (adequate to the hypothesis of a quasar or an 
X-ray binary); 
\item a Raymond spectrum with one or two temperatures (that could describe a 
supershell component);
\item a multi-color black body disk (MCBB - adequate to a black hole candidate, 
as in \citet{maki}. 
\item different combinations of the above models to check for multiple
components.
\end{itemize}
The fit results are shown in Table~\ref{models}. The quoted errors for the
spectral parameters correspond to the 90\% confidence level for 1 interesting
parameter. 
We first analysed separately the different data sets (ROSAT/PSPC, SAX/LECS+MECS,
ASCA/SIS). 
A simple power law gives a good fit for the PSPC and an acceptable fit for the 
SAX data, but with very
different photon indexes ($2.24\pm0.24$ for PSPC and $\sim 1.4$ for SAX). This
could be the signature of either strong spectral variability or the presence of
a soft component not immediately recognizable in the SAX data.  
In particular, the best fit for the MECS
data is a power law with with $\Gamma=1.40\pm 0.20$. The same simple model can 
be used to acceptably fit the composite MECS+LECS spectrum, obtaining 
$\Gamma=1.37\pm 0.18$. We also checked for the presence of a Fe K$\alpha$ line, 
by adding to the power law model a Gaussian line with
fixed energy either at E=$6.4$ keV or E=$6.7$ keV. The F-test
performed after adding the line at E=$6.4$ keV (the best choice) indicates that 
adding this component is not justified if we retain 5\% of F-test probability 
value as the confidence level for significative component addiction.

An alternative model for PSPC data is the MCBB, with a temperature of $0.60\pm
0.06$ keV. This model also gives an acceptable fit for ASCA data, with 
$T=1.24\pm 0.03$ keV, but is not adequate to fit the SAX data 
($\chi^2/\nu=73/45$).

We also made the assumption that the soft spectrum does not vary in the
time span from the set of PSPC observations to the SAX observation,
since we see no significant variability in the
PSPC hardness ratio, and a rather small flux difference between PSPC and SAX
measurements (see Figure~\ref{cntrate}). With this assumption we made a 
simultaneous fit of PSPC and MECS data, finding that an excellent fit can be
obtained with a power law with $\Gamma=1.32^{+0.22}_{-0.38}$ plus a disk 
component (i.e. an MCBB), with $T=0.29^{+0.16}_{-0.11}$ keV. The same model 
was then used to fit LECS+MECS and ASCA/SIS data. For LECS+MECS data we 
find a marginal improvement (F-test probability of 4\%) of the fit probability 
related to the reduced 
$\chi^2$, with respect to the simple power law model,
and a set of fit parameters in very good agreement with those found for 
PSPC+MECS. For ASCA/SIS data the spectrum appears dramatically 
different: a much higher temperature for the disk component (T=$1.24\pm 0.03$ 
keV) is needed than for 
the other data sets, and there is no evidence for the presence of a power law 
component.

The Raymond-Smith model (marked with 4 in Table~\ref{models}), 
gives a good fit only to the SAX data, but with a very high kT (basically
confirming the hard power law), incompatible with the soft emission that may be
expected from a superbubble \citep[e.g.][]{wangq,wangd}.
Moreover, if we consider the results obtained with this model 
for the observations made with the other instruments, we find strong variations 
both in the temperature and
in the emission measure. The latter is given as the model normalization 
(multiplied by a factor containing the distance), and we find that the three 
normalizations relevant to each instrument are not compatible with each other.
The same happens for the temperature where the one found with SAX is at least 
one order
of magnitude higher than that found with the two other data sets. As we do not
expect such a high spectral variability in a few years period from a SNR, we 
can confidently reject this model as inadequate to account coherently for all 
the spectra we observe.

Concluding, the most plausible scenario for the spectral behaviour of X-9 is
that of a composite power-law + MCBB, where the MCBB temperature and
emission measure increase with increasing source luminosity.

\section{DISCUSSION} 
\label{disc}
Although $\sim 10\%$ of the emission observed in the ROSAT HRI could be
extended, the temporal and spectral characteristics of X-9
strongly suggest that most of the emission is due to a compact object.
The strong variability (up to a factor of $\sim 4$ in a monthly timescale) 
observed in the X-9 light curve (Figure~\ref{cntrate}) excludes the hypothesis 
of Supernova emission and points to a compact X-ray source.
From the analysis of the spectral data sets, we see that this flux
variation is correlated with a spectral variation.

\begin{figure}[h]
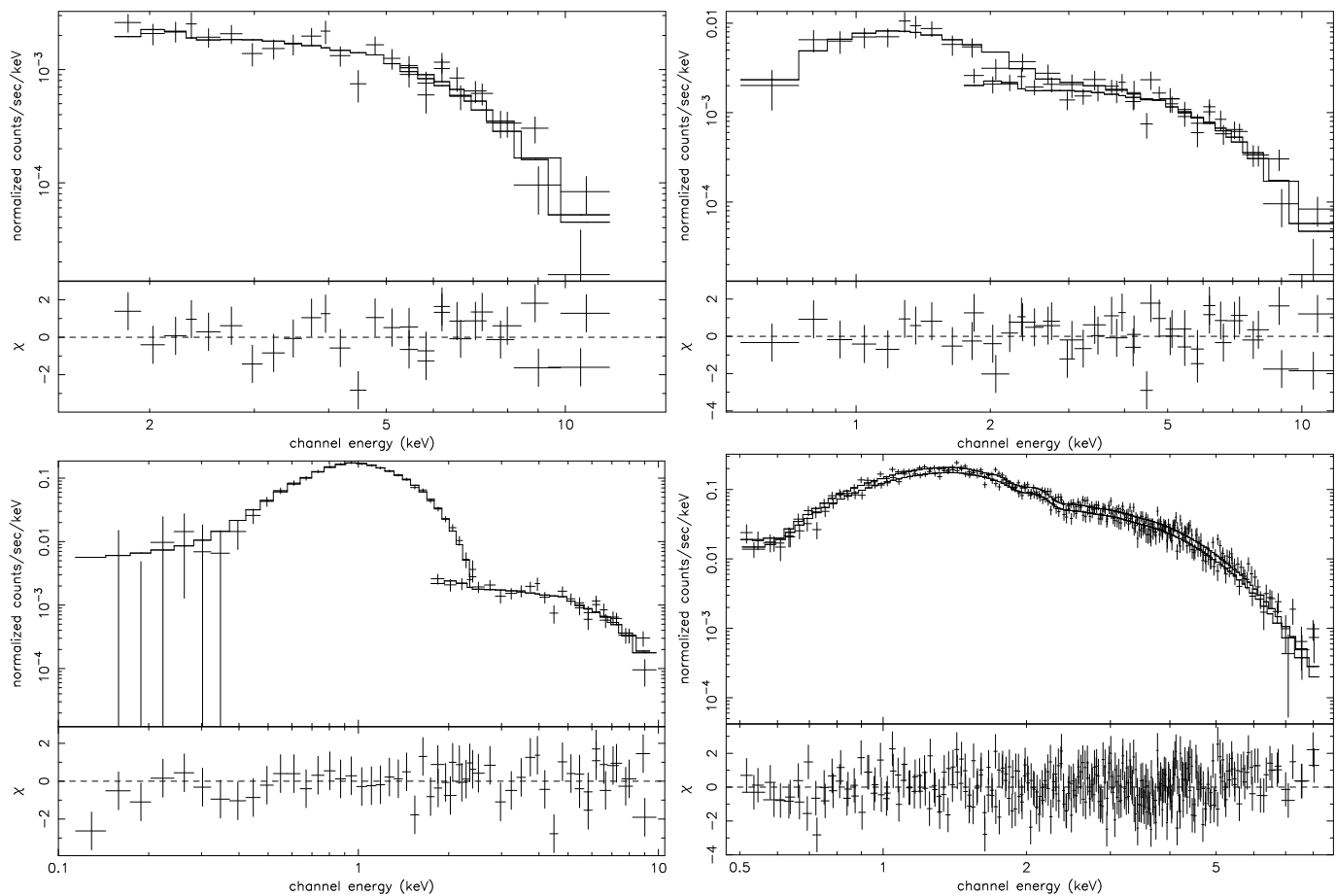

\centerline{\psfig{figure=f7a.eps,width=9cm,angle=270}
\psfig{figure=f7b.eps,width=9cm,angle=270}}
\centerline{\psfig{figure=f7c.eps,width=9cm,angle=270}
\psfig{figure=f7d.eps,width=9cm,angle=270}}
\caption{X-9 spectra, best fit models and residuals in $\chi^2$ units. 
Upper left: MECS (power-law). Upper right: MECS + LECS (power-law + MCBB).
Lower Left: PSPC + MECS (power-law + MCBB). Lower right: ASCA
(MCBB). Relevant parameters are in Table~\ref{models}. ASCA data are from the 
two SIS detectors.}
\label{bestfit}
\end{figure}

The spectral behaviour of X-9 let us conclude that this source is not a far-away
background QSO (as suggested by \citet{ezoe}, who analyzed a set of ASCA/GIS
data pointed on M81), but rather an ultra-luminous binary in Ho~IX. The best fit
temperature from the ASCA spectrum ($T\sim 1.2$ keV) is well in excess of the 
typical values we would expect from AGNs \citep[T$\lesssim 0.1$ keV for 
M$>10^6M\odot$;][]{piro}. Moreover, the BeppoSAX low state spectrum
(Figure~\ref{bestfit}) shows a very hard power law ($\Gamma\sim 1.3$), by far
harder than the $\Gamma\sim 1.7\div 2.0$ power law seen in AGN spectra 
\citet{rey}.

The spectral characteristics instead resemble those seen in some time-variable
X-ray binaries \citep[see e.g.][]{tan95} and in particular are in very good 
agreement with those found for ultra-luminous sources in spiral galaxies by 
\citet{maki}.
Under this hypothesis we are observing a "low" luminosity state with SAX and 
ROSAT ($L_{(0.5-2.4)}\sim3\times 10^{39}$), and a more luminous "high" state 
with ASCA ($L_{(0.5-2.4)}=7.6\times 10^{39}$ ergs~s$^{-1}$). According to the 
model described by \citet{esin}, the high state spectrum is dominated
by a very hot disk, while in the low state the disk is either much 
colder or not visible at all. We have marginal evidence of its presence in the
LECS+MECS spectrum, while the PSPC data clearly show a softer component that 
cannot be modelled with the simple power law used for the SAX data. As 
explained in Section~\ref{spec}, we can reasonably assume that the spectrum 
does not change much between PSPC and SAX observations, and make a simultaneous 
fit of the two data sets. The best fit is then a power law plus an MCBB 
component, whose temperature is $\sim 1/3$ of that found for ASCA data. This
result suggests that some change in the structure of the emission region may
have caused the observed change in the spectra. 

The spectrum appears
to be absorbed: the values of the $N_H$ are compatible for all spectra (ASCA,
SAX and PSPC) and the mean value (weighted mean) is $0.18\pm 0.02 \times
10^{22} cm ^{-2}$, that is $\sim 4$ times the Galactic line of sight value. This
excess could be due to the presence of a diffuse medium in the region
surrounding the source, as suggested by the HRI image (see Section~\ref{image}) 
and by the optical images \citep{mh} that show the presence of a 
H~{\sc II} region corresponding to the position of X-9. 

\subsection{The "High State"}
The "high" state is crucial to derive some physical parameters of the system.
In the hypothesis of an accreting black hole in the outer region of Holmberg IX,
X-9 would be at a distance of 3.4 Mpc \citep[from][the distance 
modulus of Holmberg IX is 27.67 and the extinction is negligible]{geo}). In the ASCA 
data there is no evidence of a power-law continuum over the disk spectrum; this 
is consistent with the hypothesis that
during the ASCA observation the disk was very close to the last stable orbit,
and that its luminosity in the ASCA energy band was very high with respect to 
the reprocessed power-law continuum. We can then infer the inner radius directly
from the normalization of the model as implemented in XSPEC. We find that,
for a distance D = 3.4 Mpc, the inner radius is $183\pm 9$ km, assuming a
face-on geometry (otherwise this number scales as $(\cos\theta)^{-1/2}$, where
$\theta$ is the inclination angle of the disk). For a Schwarzschild black hole 
the mass is related to the inner radius as:
\begin{equation}
R_{in}=\frac{6GM}{c^2}=8.86\frac{M}{M\odot} km
\end{equation}
that leads to a mass of $21\pm 1 M\odot$. The bolometric luminosity can be
evaluated from the relation between maximum disk color temperature and inner 
radius, as in \citet{maki}:
\begin{equation}
L_{bol}=4\pi(R_{in}/\xi)^2\sigma(T_{in}\kappa)^4
\end{equation}
where $\sigma$ is the Stefan-Boltzmann constant, $\kappa\sim 1.7$ is the ratio
of the color temperature to the effective temperature 
\citep["spectral hardening factor", e.g.][]{shi} and $\xi=0.412$ \citep{kubo} 
accounts for the fact that $T_{in}$ occurs at a radius somewhat larger that 
$R_{in}$. The above expression evaluated for the ASCA spectral parameters gives 
$L_{bol}=8.1\times10^{39}$; this is a factor of $\sim 3$ in excess of the 
Eddington luminosity limit for a $21\pm 1 M\odot$ black hole that is 
\begin{equation}
L_{Edd}\sim \frac{4 \pi G M m_p c}{\sigma_T}
=1.3\times10^{38} \frac{M}{M\odot} = {2.7\pm 0.1}\times10^{39} 
ergs~s^{-1}M\odot^{-1}
\end{equation}
where $M$ is the mass of the accretor, $m_p$ is the mass of the
proton, $\sigma_T$ is the Thomson cross section. Even assuming an accretion 
efficiency $\eta\sim 1$ we would need a black hole 
mass of at least $54 M\odot$ for the observed $L_{bol}$ to be compatible with 
the Eddington limit (where $L_{bol}=\eta L_{Edd}$). We then find the same 
inconsistency, in the derived physical parameter of the source reported by
\citet{mizuno} for two ultra-luminous sources in NGC 4565 and summarized 
by \citet{maki} for a larger sample of objects. \citet{maki} examine a 
number of possible solutions, and suggest that if these objects are rapidly 
spinning black holes, the high temperature can be explained by the accretion
disk getting closer to the black hole, and thus hotter.

As we do not have independent estimates of the distance of X-9, we discuss here
the possibility that the distance is somehow overestimated.
If X-9 is closer than 3.4 Mpc, we could explain the apparent 
inconsistency between the mass evaluated from
the MCBB spectral parameters and that required assuming
the Eddington limit. In fact this quantities scales as:
\begin{equation}
M_{Edd}\propto D^2 \hspace{2cm} M_X\propto R_{in}\propto D
\end{equation}
so that the ratio of these two mass estimates is
proportional to the distance. In order to make the two mass estimates agree, we
should have overestimated the distance of X-9 by a factor of at least 2.6
(assuming $\eta=1$); this 
correction would give a distance of $\sim1.3$ Mpc, and a new estimated accreting
mass of $\sim 8 M\odot$, with a luminosity of $\sim 1.0\times10^{38}$ 
ergs~s$^{-1}$. However, if we make this scenario even more realistic, we should 
take $\eta\sim 0.06$,
that is the typical value assumed for Schwarzschild black hole efficiency. In
this case we would get D=145 kpc, M=$0.88M\odot$, i.e. not a black hole anymore,
but a white dwarf. However, this results has severe drawbacks:
the location of the source would be in the intergalactic
space between the Milky Way and M81, a rather unusual location for an otherwise
tipically galactic source, and the spectrum we would expect from a
source of this kind should be much softer than the one we actually observe, 
with a temperature of the order of $10^{-2}$ keV \citep[see e.g.][]{kaha}. 

We can also exclude the possibility suggested by \citet{stocke}, from the study 
of MS~0317.7-6647, located in the vicinity of NGC1313.  
The authors suggest that this source could be inside our Galaxy, 
at the distance of $\sim 150$ pc, being an old isolated neutron star which is
no longer rotating or pulsing, but is still emitting X-rays by slow accretion of
interstellar gas onto its magnetic poles from the thick surrounding interstellar
matter. If this were the case of X-9 too, the luminosity would be of the order
of $10^{31}$. However, if the spectrum observed with ROSAT is in the
temperature range predicted for this kind of sources \citep{colpi}, the
ASCA spectrum is much too hard. Moreover, the presence of the power law
detected with SAX/MECS cannot be justified by this model, and the same holds for
the spectral variability we observe between SAX and ASCA observations. 

If we retain the location of the source to be in 
Holmberg IX (i.e. at a distance of 3.4 Mpc), we can follow the argumentation put
forward by \citet{maki}, who make the hypothesis that the accreting
object is a Kerr black hole. In this case the disk can come closer to the
central mass, and the last stable orbit can go down to 0.5 Schwarzschild radii,
implying that the temperature can get higher, even if we neglect any strong 
relativistic effects. To apply this simple correction to our results, we remind
that the last stable orbit radius of a rotating black hole is related to the
mass as 
\begin{equation}
R_{in}=\alpha\frac{6GM}{c^2}=8.86\alpha\frac{M}{M\odot} km
\end{equation}
where $1/6<\alpha<1$. In the extreme case of $\alpha=1/6$, corresponding to a
maximally rotating black hole, we get
$M=126M\odot$. This mass implies $L_{Edd}=1.9\times10^{40}$, that agrees well 
with the measured bolometric luminosity $L_{bol}$, if we  
assume a radiation efficiency of $\eta\sim 0.4$ ($L_{bol}=\eta L_{Edd}$), as 
can be reached by a maximally
rotating black hole, fed with a prograde disk \citep[see, e.g,][]{sun}.

Therefore, while the model of a non-rotating black hole leads to severe
drawbacks, the hypothesis of a rotating black hole, along the line suggested by
\citet{maki}, may give a more self-consistent scenario, leading to  a
$126M\odot$ black hole. It is worth noting that the existence of black holes of 
this mass is still quite controversial, being much smaller than galactic 
nuclear black holes ($10^7 - 10^8 M\odot$) and much larger than
stellar sizes black holes ($10 M\odot$), thus suggesting new interesting 
scenarios to massive 
black holes formation. \citet{madau} suggested that such objects, with masses
$\gtrsim 150 M\odot$ can be the end product of pre-galactic star formation
episodes: their progenitors would be zero-metallicity (Population III) very
massive stars.
The high mass limit would be relaxed to smaller values if X-9 is a beamed
source. Beaming in ultra-luminous galactic sources has recently been advanced as
an alternative scenario to the very massive black hole hypothesis \citep{king}

\subsection{The "Low" State}
The nearly constant PSPC hardness ratios and the good 
agreement between PSPC and LECS spectra suggest that no significant 
spectral changes have occured in the soft X-ray emission from X-9 during the    
"low state" observations. 
The PSPC and SAX(MECS+LECS) spectra
can be well described with a MCBB model with temperature T$\sim 0.28$~keV plus a
power-law with photon index $\Gamma \sim 1.35$. 

The observed spectral features resemble those observed in the low state of 
galactic black hole 
candidate, such as the well studied Cyg X-1 \citep[see, e.g.][]{gierl,fron}, 
Nova Muscae \citep{esin}, 
or Sgr~A$^*$ \citep{nara}. These sources are observed to
spend most of their time in a low intensity state, characterized 
by a power law continuum and, for high enough accretion rates, a soft thermal
component arising from the accretion disk. In most cases there is
also evidence of fluorescent emission from Fe, with the detection
of a line at 6.4 keV.  These features are interpreted in the framework 
of the ADAF \citep[Advection Dominated Accretion Flow, see, e.g,][]{esin} 
model as the signature of a geometry where the disk is truncated far
from the last stable orbit, and in the region between the disk and the compact
object (the ADAF region) a substantial part of the energy is stored in the 
accreting gas rather than being radiated. When the accretion rate grows over a 
critical value, the ADAF disappears and the disk approaches the last stable
orbit. In this high state the emission is dominated by the disk thermal
spectrum, in a way similar to what we observe in the ASCA spectrum of this
source. 

However the power law index predicted by the ADAF model (and observed in the
sources above -- $\Gamma\gtrsim 1.5$) for the low state is larger than the 
index we measure in our 
data, suggesting that this model may not be entirely adequate to describe our 
source. A behaviour  
closer to that of X-9 was observed with ASCA in 1E~1740.7-2942, a bright 
source near to the Galactic center \citep{saka,sidoli}. 
The broad band spectrum of this source strongly suggests that it 
is indeed a stellar mass black hole. 
\citet{saka} find that the 2.0-10.0 keV spectrum can be described
with a power law of photon index $\Gamma$ in the range 0.9-1.3. They do not 
find any evidence of thermal component, but this could be due to the harder 
energy band they use in their analysis (2.0-10.0 keV vs. 0.2-10.0 keV of our 
data). This source, indeed, shows no evidence for a line at 6.4 keV. 

In summary, the low state spectrum of X-9 resembles closely the 
spectra of other black hole candidates in their low state, with the only
difference of a slightly flatter power law index. 

\section{CONCLUSIONS}
\label{final}
We have analysed all the available X-ray observations of the M81 region in 
order to understand the nature of the ultra-luminous X-ray source M81 X-9. Our 
results can be summarized as follows:
\begin{enumerate}
\item The X-ray source is a point-like object, possibly surrounded 
by a faint diffuse medium, that could account for about 10\% of the total 
emission. This result needs a more accurate verification, that can 
only come from observations with high resolution X-ray cameras, like those
on board of the Chandra Observatory. The optical counterpart appears to be a 
faint blue object, embedded in an H$\alpha$ cloud. The nature of this
object is still unclear, and optical observations are also needed to clarify 
this point.
\item The flux of X-9 is strongly variable and this variability points to 
a compact object. This source was observed over twenty years with 
{\it Einstein}, ROSAT, ASCA and BeppoSAX, with almost countinous coverage 
during the last eight years. The light curves show 
non periodic flux variations of up to a factor of 4.
\item The ROSAT/PSPC and SAX data were all collected during low
intensity states ($L_X\sim 4\times 10^{39}$ erg~s$^{-1}$) and have consistent 
spectral distributions, while the high state ASCA spectrum ($L_X\sim 
8\times 10^{39}$ erg~s$^{-1}$) is significantly different. 
The observed spectrum can be well fit with a MCBB model with $T\sim 1.24$ keV 
for the high state ASCA data and with a colder ($T\sim 0.3$keV) MCBB plus a 
flat ($\Gamma \sim 1.3$) power-law for the PSPC and SAX data. 
\item The intensity and spectral variability of X-9 supports the scenario
of an accreting black hole binary. From the
ASCA best fit parameters we infer a Kerr black hole mass of $\sim 126 M\odot$, 
if at the distance of 3.4 Mpc and assuming non-beamed X-ray emission.
\end{enumerate}

We thank Andreas Zezas for discussions on ultra-luminous sources in galaxies, 
Sam Dyson for help in the initial ROSAT data reduction, Rosario Iaria for 
discussions on accretion processes. This work was supported in
part by NASA grant NAG5-2946 and NASA contract NAS8-39073(CXC) and in part by
MURST. This research has made use of the HEASARC online database and of the ESO
online DSS.

\newpage
\begin{deluxetable}{crrrrr}
\tablecaption{Log of Einstein, ROSAT, SAX, ASCA observation of X-9.\label{tbl-1}}
\tablewidth{0pt}
\tablehead{
\colhead{Instr.}     & \colhead{Ref.}  &
\colhead{Sequence nr.}   & \colhead{Live Time}           &
\colhead{Start Date} & \colhead{Angle\tablenotemark{a}}}
\startdata
{\it Einstein}  & E-IPC & 2102      &    6515&27/04/79& 11.7'\\
                & E-HRI &  585      &   19627&27/09/79& 12.3'\\ \tableline
ROSAT PSPC& 1P  &rp600101a00&    9296&25/03/91& 12.7'\\
          & 2P  &rp600110a00&   12717&27/03/91& 38.6'\\
          & 3P  &rp600052n00&    6588&18/04/91& 31.1'\\
          & 4P  &rp600110a01&   12238&15/10/91& 38.6'\\
          & 5P  &rp600101a01&   11085&16/10/91& 12.7'\\
          & 6P  &rp600382n00&   27120&29/09/92& 12.7'\\
          & 7P  &rp180015n00&   17938&03/04/93& 13.5'\\
          & 8P  &rp180015a01&    8731&04/05/93& 12.5'\\
          & 9P  &wp600576n00&   16412&29/09/93& 19.5'\\
          &10P  &rp180035n00&   17800&01/11/93& 12.5'\\
          &11P  &rp180035a01&    4234&07/11/93& 12.5'\\
          &12P  &rp180050n00&    1849&31/03/94& 12.5'\\ \tableline
ROSAT HRI & 1H  &rh600247n00&   26320&23/10/92& 12.5'\\ 
          & 2H \tablenotemark{b} &rh180015n00&    1688&16/04/93& 12.5'\\ 
          & 3H  &rh600247a01&   21071&17/04/93& 12.5'\\ 
          & 4H  &rh600739n00&   19902&19/10/94& 12.5'\\ 
          & 5H  &rh600740n00&   18984&13/04/95& 12.5'\\ 
          & 6H  &rh600881n00&   14826&12/10/95& 12.5'\\
          & 7H  &rh600882n00&   18328&15/04/96& 12.5'\\
          & 8H  &rh600882a01&    5091&27/10/96& 12.5'\\ 
          & 9H  &rh601001n00&   19231&29/03/97& 12.5'\\
          &10H  &rh601002n00&   12590&30/09/97& 12.5'\\ 
          &11H  &rh601095n00&   12590&25/03/98& 12.5'\\ \tableline
BeppoSAX  &MECS &40732001   &  100287&04/06/98&  13' \\ 
          &LECS &           &   43931&        &      \\ \tableline
ASCA      &SIS  &57048000   &   33000&06/04/99&  5'  \\ 
\enddata
\tablenotetext{a}{Off axis angle of the source on the detector}
\tablenotetext{b}{Sequence 2H was not used (see text)}
\label{log} 
\end{deluxetable}

\begin{deluxetable}{r c c c c c c c c c c c c c}
\tablecaption{Variability tests on individual sequences. \label{tbl-2}}
\tablewidth{0pt}
\tablehead{
\colhead{Seq.ref}&\colhead{KS test}&\colhead{CvM test}&
\colhead{Collura}&\colhead{Var.}

}
\startdata
E-IPC& NO & NO & NO &   \\
E-HRI& NO &95\%& NO &   \\
1P   & NO &NO  &90\%&   \\
2P   & -  & -  & -  &   \\
3P   & -  & -  &95\%&   \\
4P   & -  & -  & -  &   \\
5P   & NO &NO  &95\%&   \\
6P   &99\%&95\%&99\%&$\times$\\
7P   &99\%&99\%&99\%&$\times$\\
8P   & NO & NO & NO &   \\
9P   & -  & -  & -  &   \\
10P  & NO &95\%&99\%&$\times$\\
11P  & NO & NO & NO &   \\
12P  & NO & NO &95\%&   \\
1H   & NO & NO &90\%&   \\
3H   &99\%&95\%&90\%&$\times$\\
4H   & NO &NO  &NO\%&   \\
5H   & NO &NO  &90\%&   \\
6H   & NO & NO & NO &   \\
7H   & NO &NO  &90\%&   \\
8H   & NO & NO & NO &   \\
9H   & NO & NO &  - &   \\
10H  & NO & NO &  - &   \\
11H  &90\%&90\%&90\%&$\times$\\
SAX  & NO & -  & -  &   \\
SIS  & NO & NO & -  &   \\

 \enddata
\tablecomments{For those observations found variable with more than 90\% 
probability, the variability 
confidence level is quoted. NO means no variability found; '-' means that 
the test could not be applied because of strong background variability or
unsufficient statistics. Variable observations are flagged with $\times$ on 
the fourth column.}
\label{kstest}
\end{deluxetable}

\begin{deluxetable}{r c c }
\tablecaption{Definition for Hardness Ratio bands. \label{tbl-3}}
\tablewidth{0pt}
\tablehead{
\colhead{}&\colhead{Soft band}&\colhead{Hard band}
}
\startdata
Whole band (HR1) & 0.11-0.42 KeV  & 0.52-2.02 KeV\\ 
Hard band  (HR2) & 0.52-0.91 KeV  & 0.91-2.02 KeV\\ 
 \enddata
\label{hrdef}
\end{deluxetable}

\begin{deluxetable}{r l c c c c c c}
\tablecaption{Spectral analysis: fit results. \label{tbl-4}}
\tablewidth{0pt}
\tablehead{
\colhead{Model}&\colhead{Instruments}
&\colhead{nH($\times10^{22}$)\tablenotemark{a}}&
\colhead{$\Gamma$\tablenotemark{b}}&\colhead{T (keV)\tablenotemark{c}}
&\colhead{$\chi^2_{\nu} (Prob)$\tablenotemark{d}}
&\colhead{$\nu$\tablenotemark{e}}
}
\startdata
1&MECS        &$0.04^{+0.6} $       &$1.40^{+0.19}_{-0.22}$ & - 
    & 1.48 (0.05)& 27\\
2&MECS        &$0.04^{+0.6}      $  &$1.42\pm 0.16        $& -
    & 1.38 (0.10)& 26\\  \tableline  

1&MECS+LECS   &$0.08^{+0.12}_{-0.04}$&$1.37^{+0.18}_{-0.17}$& -
    & 1.04 (0.38)&44 \\ 
3&MECS+LECS&$0.3^{+0.5}_{-0.3}$&$1.28^{+0.28}_{-0.36}$&$0.28^{+0.40}_{-0.06}$
    & 1.20 (0.18)&42 \\    
4&MECS+LECS&$0.06^{+0.10}_{-0.02}$& -                 &$28^{+50}_{-14}$
    & 1.19 (0.18)&44 \\ 
5&MECS+LECS&$0.04^{+0.03}        $& -                 &$3.0^{+0.7}_{-0.5}$ 
    & 1.62 (0.00)&45\\  \tableline

1&PSPC        &$0.22\pm0.05         $&$2.24\pm0.24$         & -  
    & 0.71 (0.88)&29 \\ 
4&PSPC        &$0.085\pm0.010       $& -         & $4.8^{+1.2}_{-0.7}$ 
    & 1.42 (0.06)&28 \\
5&PSPC        &$0.10^{+0.03}_{-0.02}$& -                    &$0.60^{+0.06}_{-0.07}$ 
    & 0.89 (0.63)&29 \\  \tableline
    
3&MECS+PSPC   &$0.18^{+0.13}_{-0.07}$&$1.32^{+0.22}_{-0.38}$&$0.29^{+0.16}_{-0.11}$ 
    & 1.04 (0.40)&56 \\   \tableline 

1&ASCA SIS    &$0.48^{+0.02}_{-0.03}$&$2.15\pm0.04         $         & -  
    & 1.53 (0.00)&318\\ 
4&ASCA SIS    &$1.01^{+0.19}_{-0.04}$&$2.58^{+0.07}_{-0.08}$&$0.119^{+0.018}_{-0.008}$
    & 1.25 (0.00)&315\\
5&ASCA SIS    &$0.177\pm0.015$       &-                     &$1.24\pm0.03$
    & 1.13 (0.05)&318\\
\enddata
\tablecomments{The first column contains the
model description according to the following code: 1) Power-law; 2)
Power-law + Gaussian line (with energy fixed at 6.4 keV) ; 3) Power-law + MCBB; 
4) Raymond-Smith thermal model (+ power law where needed); 5) MCBB. }
\tablenotetext{a}{Column density. The lower limit is set to $4.1\times 10^{20}$
cm$^{-2}$, i.e. the galactic line of sight value.}
\tablenotetext{b}{Power-law photon index. }
\tablenotetext{c}{Temperature of the MCBB or of the Raymond spectrum, whichever of 
these models have been used.}
\tablenotetext{d}{Reduced $\chi^2_{\nu}$ and probability value}
\tablenotetext{e}{Number of degrees of freedom}
\label{models}
\end{deluxetable}

\clearpage

\clearpage

\clearpage

\clearpage

\clearpage

\clearpage

\clearpage

\end{document}